\documentclass[12pt]{revtex4}
\usepackage{graphicx}
\usepackage{dcolumn}
\usepackage{bm}
\begin{document}

\title{EPR before EPR: a 1930 Einstein-Bohr thought experiment revisited}

\author{Hrvoje Nikoli\'c}
\affiliation{Theoretical Physics Division, Rudjer Bo\v{s}kovi\'{c}
Institute,
P.O.B. 180, HR-10002 Zagreb, Croatia.}
\email{hrvoje@thphys.irb.hr}

\date{\today}

\begin{abstract}
In 1930 Einstein argued against consistency of the time-energy uncertainty 
relation by discussing a thought experiment involving a measurement of 
mass of the box which emitted a photon. Bohr seemingly triumphed over Einstein
by arguing that the Einstein's own general theory of relativity 
saves the consistency of quantum mechanics. We revisit this thought
experiment from a modern point of view 
at a level suitable for undergraduate readership
and find that neither Einstein nor
Bohr was right. Instead, this thought experiment should be thought of as
an early example of a system demonstrating nonlocal ``EPR'' quantum correlations, 
five years before the famous Einstein-Podolsky-Rosen paper. 
\end{abstract}

\maketitle

\section{Introduction}

The nonlocal nature of quantum correlations in systems with two or more particles
is today a well understood and widely known \cite{bell,merm1,merm2,jord,merm3,laloe,nik-myth,scholarpedia} property of quantum mechanics (QM).
Even though the nonlocal nature of quantum correlations was first clearly recognized by Bell
in 1964 \cite{bell-1964} and experimentally verified by Freedman and Clauser in 
1972 \cite{EPR-test}, 
these nonlocal correlations are widely known as EPR correlations,
after a 1935 paper by Einstein, Podolsky, and Rosen \cite{EPR}.
Nevertheless, the nonlocality of QM has not been even conjectured
in the EPR paper. Instead, the paper was an attempt to prove
that QM was incomplete,
where one of the assumptions in the proof was that nature is local.
Today we know that Einstein, Podolsky, and Rosen were wrong,
in the sense that this assumption of locality 
was wrong. It does not imply that QM {\em is} complete (at the moment there
is no consensus among physicists on that issue \cite{nik-myth}), 
but today the EPR argument against completeness is no longer appropriate.

Even though nonlocality of QM has not been discovered in the famous 1935 EPR paper,
that paper played a pivotal historical role by influencing Bell to discover nonlocality
in 1964 \cite{bell-1964}. 
(To be more precise, the Bell's ``discovery"
was a purely theoretical result, not an experimental one.
He studied general 
properties of local theories describing measurement outcomes
in terms of objective properties possessed by the system. 
As a result he (i) derived an inequality involving 
correlations between measurement outcomes that any local theory 
of that kind should obey, and (ii) found out that QM predicts
violation of that inequality.)
For that reason, it seems to be widely believed that
the EPR paper is the first historical example of a quantum system which cannot be understood correctly without invoking quantum nonlocality. 

In the present paper we point out that the EPR paper is not the first such example.
Instead, a very similar example was provided by Einstein alone in 1930,
i.e., five years {\em before} the EPR paper. In fact, this Einstein's example from 1930,
together with the Bohr's reply, is well known and reviewed in many books
\cite{jammer,wheeler,greenstein,aharonov}. In this example, Einstein 
presents a paradox in QM suggesting that QM is inconsistent, while Bohr attempts to save consistency of QM by combining QM with the Einstein's general theory of relativity.
Yet, despite the historical importance of that paradox, and despite the similarity
with the EPR paradox, and despite the fact that this paradox is often presented in the books,
a correct resolution of it is rarely presented in the literature. Instead, many books like \cite{wheeler,greenstein,aharonov}
still present the Bohr's resolution of the paradox in terms of general relativity
as if such a resolution was correct.

Jammer in his book \cite{jammer} is less superficial on that 
issue by discussing several criticisms of the Bohr's resolution, and yet none of these criticisms
includes the correct resolution based on quantum nonlocality.
At another place in \cite{jammer} 
Jammer comes very close to the correct resolution through the words:
\begin{quote}
Einstein's own narrative shows in detail
how the idea of a gravitating photon box gradually changed into that of a
system of two interacting particles, as later used in the Einstein-Podolsky-Rosen paradox. 
... The final
elimination of the box itself and its replacement by a second ``particle"
may have been prompted by the following development.
\end{quote}

Some critical discussions of the Bohr's resolution of the paradox have also been
presented after the Jammer's book in the papers \cite{unruh,torre1,hnizdo1,torre2,hnizdo2,dieks}.
Most of these papers also do not present explicitly the correct resolution of the paradox in terms of quantum nonlocality, but some of them \cite{torre1,torre2} indicate such a resolution through some brief side remarks similar to those of Jammer above. The only paper we are aware that explicitly presents
the correct resolution in terms of quantum nonlocality is \cite{dieks}.

Unfortunately, these words by Jammer, brief side remarks in \cite{torre1,torre2},
and even the explicit claims in  \cite{dieks},
remained widely unnoticed in the physics community
and the correct resolution of the Einstein's 1930 paradox remained widely unrecognized. 
Thus, we believe that it is both important for historical reasons and instructive 
for pedagogic purposes to present {\em in detail} the correct resolution of that Einstein 1930 
paradox. 
Hence, this is what we do in the present paper. 

The paper is organized as follows. We first review 
the Einstein 1930 paradox and the Bohr's resolution as presented in the existing
literature \cite{jammer,wheeler,greenstein,aharonov}, in Sec.~\ref{SEC2}. 
After that, in Sec.~\ref{SEC3} we revisit this paradox from a modern point of view and find
that the key to the correct resolution of the paradox is quantum nonlocality
(and not general relativity as suggested by Bohr in 1930).
Finally, the conclusions are drawn in Sec.~\ref{SEC4}.

The paper is written at a level suitable for undergraduate readership
and contains material useful for physics education at university level.
For example, it contains some simple qualitative explanations of the nonlocal
features of QM including the wave-function collapse, as well as a quantitative explanation of the time-energy 
uncertainty relation presented in the form that can rarely be found in textbooks.

\section{Einstein's 1930 paradox and Bohr's resolution}
\label{SEC2}

\subsection{Einstein's paradox} 

At the Sixth Solvay conference in 1930 Einstein attempted to challenge
the consistency of the time-energy uncertainty relation
\begin{equation}\label{e1}
\Delta E \, \Delta t \gtrsim \hbar .
\end{equation}
For that purpose he considered the following thought experiment. Consider a box filled with photons
open during an arbitrarily short time $\Delta t$. During this time a photon is emitted,
so one knows the time of emission with an arbitrary precision $\Delta t$. If 
(\ref{e1}) is correct, then a low time uncertainty $\Delta t$ implies a high uncertainty of the photon energy
\begin{equation}\label{e2}
\Delta E \gtrsim \frac{\hbar}{\Delta t} .
\end{equation}
However, despite the arbitrarily small $\Delta t$, the energy of the photon can be measured
with an arbitrary precision as well, in the following way. One can measure the mass $m$ 
of the box before and after the photon left the box. (For concreteness, one can do that
by reading the position of the box hanging on a spring scale in the gravitational field, which determines
the weight of the box.) From the change of the mass and $E=mc^2$ one can determine
the change of energy of the box with arbitrary precision. But total energy must be conserved,
so the energy decrease of the box must be equal to the energy carried by the photon.
In this way, by measuring the mass of the box with arbitrary precision one also determines
the energy of the photon by arbitrary precision, in contradiction with (\ref{e2}).

\subsection{Bohr's resolution}

At first Bohr didn't know how to reply to the Einstein's paradox, 
but eventually he constructed a reply which we now review.
For the sake of clarity we split his response into three steps.

The first step considers the precision of the measurement of energy. If energy of the box is measured
with the precision $\Delta E$, then $E=mc^2$ implies that mass is measured with the precision
$\Delta m= \Delta E/c^2$. This implies that the gravitational force $F=mg$ is determined by the
precision
\begin{equation}
 \Delta F= \Delta m \, g = \Delta E \frac{g}{c^2}.
\end{equation}
The force acting during the time $t$ determines the momentum transfered 
from the gravitational field
 to the box. This momentum is $p=Ft$, 
so the transfered momentum is determined by the precision 
\begin{equation}
\Delta p=\Delta F\, t = \Delta E \frac{gt}{c^2}.
\end{equation}
Therefore, the energy of the box is measured by the precision
\begin{equation}\label{b1}
 \Delta E = \Delta p \frac{c^2}{gt}.
\end{equation}

This does not yet contradict the Einstein's conclusion that energy can be measured with arbitrary
precision, but the second step considers another effect. Einstein argued that time of the photon emission 
can be measured with arbitrary precision, but he has not taken into account 
general relativity which says that there is a relation between time and position.
Namely, on Earth there is a weak gravitational potential $\phi(x) \approx gx+{\rm const}$, 
so general relativity for weak gravitational
fields says that the lapse of time depends on the position $x$ as
\begin{equation}\label{time}
 t=\left( 1+\frac{\phi(x)}{c^2} \right) t_0 \approx \left( 1+\frac{gx+{\rm const}}{c^2} \right) t_0 ,
\end{equation}
where $t_0$ is time lapsed at the position $x_0$ at which $\phi(x_0)=0$. 
It is convenient to choose the additive constant 
in (\ref{time}) so that $x_0$ is the position corresponding to the Earth's surface.
This implies that there is uncertainty in time of photon emission from the box
due to the uncertainty of the box position
\begin{equation}\label{b2}
 \Delta t \approx \frac{g \Delta x}{c^2} t_0 .
\end{equation}

The third step is to consider the product $\Delta E \, \Delta t$. Eqs. (\ref{b1}) and (\ref{b2}) give
\begin{equation}
 \Delta E \, \Delta t \approx \frac{t_0}{t} \Delta p \, \Delta x \approx \Delta p \, \Delta x ,
\end{equation}
where we have used $t_0/t \approx 1$. 
(The relation $t_0/t \approx 1$ means that the 
time lapsed at $x_0$ does not differ much
from the corresponding time lapsed at $x$.
This is because 
the effects of general relativity are small on Earth, i.e., 
$\phi(x)/c^2\ll 1$ in (\ref{time}).)
So far the analysis was classical, but at this step
it is legitimate to use the quantum bound $\Delta p \, \Delta x \gtrsim \hbar$, because 
Einstein in his argument has not refuted the momentum-position uncertainty relation. This finally gives
\begin{equation}\label{b3}
 \Delta E \, \Delta t \gtrsim \hbar ,
\end{equation}
in contradiction with the Einstein's conclusion. The crucial step in the derivation of (\ref{b3}),
which Einstein did not take into account, was general relativity, i.e., Eq.~(\ref{time})).

In summary, to defeat Einstein, Bohr has used the Einstein's own weapon - general relativity.
For that reason it is often considered to be the greatest triumph of Bohr over Einstein.
After that point, Einstein stopped trying to prove the inconsistency of QM, 
and instead tried to prove that QM is incomplete.

\section{The paradox revisited from a modern point of view}
\label{SEC3}

From a modern perspective, the Bohr's reply to Einstein does not look very rigorous.
In particular, it does not involve any reference to a state in the Hilbert space.
Besides, the Bohr's reply suggests that QM cannot be made self-consistent 
without incorporating general relativity into it, which is very far from a modern view of QM.
(In fact, a consistent incorporation of general relativity into QM is one of the greatest
unsolved problems in modern physics.) This motivates us to revisit 
the Einstein thought experiment from a different, more modern perspective.
(By ``modern'', we mean modern compared to the understanding
of QM in 1930.)
The discussion we present is quite elementary, but
in the context of the Einstein-Bohr 1930 thought experiment it is widely unknown.

\subsection{Energy and time uncertainties as properties of the quantum state} 

Let us first discuss what exactly does it mean that the photon's energy is uncertain.
In terms of states in the Hilbert space, it means that the quantum state is a superposition of
different energies $E=\hbar \omega$, where $\omega$ is the photon's frequency.
For example, if the photon is in a pure state represented by a unit vector in the
Hilbert space, then this state can be written as
\begin{equation}\label{psi}
 |\psi\rangle = \int d\omega \, c(\omega) |\omega\rangle ,
\end{equation}
where, for simplicity, the spin degrees of freedom of the photon are suppressed.
The state is normalized such that $\langle\psi|\psi\rangle = 1$, which implies that
\begin{equation}
\int d\omega \, |c(\omega)|^2 =1 ,
\end{equation}
and $|c(\omega)|^2$ is the probability density that the photon has the frequency $\omega$.
The uncertainty of frequency $\Delta\omega$ is given by
\begin{equation}\label{e12}
 (\Delta\omega)^2 = \langle \omega^2 \rangle - \langle  \omega \rangle^2 ,
\end{equation}
where 
\begin{equation}\label{e13}
 \langle \omega^n \rangle =\int d\omega \, |c(\omega)|^2 \omega^n .
\end{equation}

One way to define the time-uncertainty $\Delta t$ 
is in terms of Fourier transforms as follows.
One first introduces the Fourier transform
\begin{equation}\label{e14}
 \tilde{c}(t)\equiv \int \frac{d\omega}{\sqrt{2\pi}} c(\omega) e^{-i\omega t} ,
\end{equation}
and then in analogy with (\ref{e12}) and  (\ref{e13}) defines
\begin{equation}\label{e15}
 \langle t^n \rangle \equiv \int dt \, |\tilde{c}(t)|^2 t^n ,
\end{equation}
\begin{equation}\label{e16}
 (\Delta t)^2 \equiv \langle t^2 \rangle - \langle  t \rangle^2 .
\end{equation}
With such mathematical definitions, in the theory of Fourier transforms 
it can be rigorously proven that
\begin{equation}\label{e17}
 \Delta\omega \, \Delta t \geq \frac{1}{2} .
\end{equation}
Thus, the quantum relation $E=\hbar \omega$ implies the uncertainty relation
\begin{equation}\label{e18}
 \Delta E \, \Delta t \geq \frac{\hbar}{2} .
\end{equation} 

Another, more physical (but mathematically less rigorous) way to define $\Delta t$ in QM
is to study the behavior of a physical clock. In this approach one thinks of $t$
as an abstract parameter appearing in our {\em theoretical}
description of dynamics, while the quantity which we really observe
is an observable $Q$, the time-dependence $Q(t)$ of which 
is described by the theory. Thus, the actual measurement of $Q$ 
can be interpreted as a measurement of abstract time $t$, so that $Q$
can be interpreted as the clock observable. 
In particular, the uncertainty $\Delta t$ can be expressed in terms of $\Delta Q$ 
through the relation
\begin{equation}\label{clock1}
 \Delta Q \approx \Delta t \left| \frac{d\langle Q \rangle}{dt} \right| ,
\end{equation}
where $\langle Q \rangle \equiv \langle \psi | \hat{Q} |\psi \rangle$ is the 
quantum-mechanical average of the observable $Q$.
The time-derivative of any quantum operator $\hat{Q}$ with only implicit dependence on time
is given by 
\begin{equation}
\frac{d\hat{Q}}{dt}=\frac{i}{\hbar} [\hat{Q},\hat{H}] ,
\end{equation}
where $H$ is the Hamiltonian. Therefore (\ref{clock1}) can be written as
\begin{equation}\label{clock3} 
\Delta Q \approx \frac{\Delta t}{\hbar} |\langle [\hat{Q},\hat{H}] \rangle | . 
\end{equation}
For any two operators $\hat{Q}$ and $\hat{H}$ it can be rigorously shown 
(see, e.g., Ref.~\cite{bal}) that
\begin{equation}\label{clock4}
 \Delta Q \, \Delta H \geq \frac{1}{2} |\langle [\hat{Q},\hat{H}] \rangle | ,
\end{equation}
so (\ref{clock3}) and (\ref{clock4}) together give the time-energy uncertainty relation
\begin{equation}\label{clock5}
 \Delta E \, \Delta t \gtrsim \frac{\hbar}{2} ,
\end{equation}
where we have identified the Hamiltonian $H$ with the energy $E$.

\subsection{Einstein's thought experiment in terms of quantum states}

Eq.~(\ref{psi}) would be an appropriate description of the state of photon 
if the photon was independent on the state of box. However this is not the case
in the Einstein's thought experiment, because
the photon is emitted by the box. The total Hamiltonian is an exactly conserved
quantity in QM, so the sum of photon energy and box energy must be conserved exactly.

For simplicity, let as assume that the initial energy $m_0c^2$ of the box 
before the emission is not uncertain.
Hence, if the photon energy after the emission is equal to $\hbar\omega$, then exact energy
conservation implies that the box after the emission must have energy $m_0c^2-\hbar\omega$.
Thus, instead of (\ref{psi}), the total joint quantum state of box and photon after the emission
must have the form
\begin{equation}\label{Psi}
 |\Psi\rangle = \int d\omega \, c(\omega) \left|\frac{m_0c^2}{\hbar}-\omega\right\rangle 
|\omega\rangle ,
\end{equation}
where $\left|\frac{m_0c^2}{\hbar}-\omega\right\rangle$ are energy-eigenstates of the box
and $|\omega\rangle$ are energy eigenstates of the photon.
We see that energy of the photon in (\ref{Psi}) is uncertain just as in  (\ref{psi}). 
Likewise, the energy of the box in (\ref{Psi}) is uncertain as much as the energy
of the photon. And yet, the sum of energies of box and photon is not uncertain at all,
because each product state $\left|\frac{m_0c^2}{\hbar}-\omega\right\rangle |\omega\rangle$
has the same energy $(m_0c^2-\hbar\omega)+\hbar\omega=m_0c^2$ not depending
on the value of the integration parameter $\omega$.
In modern language, (\ref{Psi}) describes the {\em entanglement} between 
box and photon, so that the photon energy is correlated with the box energy.

Now let us consider the measurement of energy.
Einstein was right that, in principle,
energy of the box can be measured with arbitrary precision and that such a 
measurement determines also the photon energy with arbitrary precision.
But what happens when we perform such a measurement? The state (\ref{Psi})
performs a transition to a new state
\begin{equation}\label{coll}
 |\Psi\rangle \rightarrow \left|\frac{m_0c^2}{\hbar}-\omega_{\rm meas}\right\rangle 
|\omega_{\rm meas}\rangle ,
\end{equation}
where $m_0c^2-\hbar\omega_{\rm meas}$ is the the energy of the box obtained by the measurement.
The transition (\ref{coll}) is often referrred to as collapse of the wave function.
But this means that {\em the measurement of the box modifies not only the state of
the box, but also the state of the photon}. It has two important consequences,
which we discuss next.

\subsection{Consistency of the time-energy uncertainty relation}

The first important consequence of (\ref{coll}) is explanation of 
how the time-energy uncertainty relation
is saved. Before the measurement, the uncertainty of the photon energy
$\Delta E$ is given by the function $c(\omega)$, through the relations
(\ref{e12}), (\ref{e13}), and $\Delta E= \hbar\Delta \omega$. Likewise,
the time uncertainty before the measurement is also given by
$c(\omega)$, through the relations (\ref{e14})-(\ref{e16}). Therefore,
(\ref{e18}) is valid before the measurement. After the measurement
the state collapses as in (\ref{coll}), which means that $c(\omega)$ collapses
into a new function
\begin{equation}\label{coll2}
 c(\omega)\rightarrow \delta(\omega-\omega_{\rm meas}) .
\end{equation}
But (\ref{e18}) is valid for {\em any} function $c(\omega)$, including 
the distribution $\delta(\omega-\omega_{\rm meas})$ on the right-hand side of (\ref{coll2}).
(There are some technical subtleties related to a proof that 
(\ref{e17}) is valid for the distribution $\delta(\omega-\omega_{\rm meas})$, but the details
are not important for our discussion. For our purpose, it is sufficient to say
that $\delta$-distribution can be regularized by a finite-width Gaussian, which allows
to study the limit in which the Gaussian width approaches zero.)
Therefore, both $\Delta E$ and $\Delta t$ change by the measurement, such that 
the property (\ref{e18}) remains intact. 

Alternatively, if $\Delta t$ is defined in terms of time-measurement by a clock as in 
Eqs.~(\ref{clock1})-(\ref{clock5}), then the time-energy uncertainty relation is saved
in the following way. The uncertainty $\Delta t$ refers to a measurement
during which the energy is measured, i.e., during which the collapse (\ref{coll})
takes place. Since $\Delta E$ is zero (or very small in a more realistic case) 
due to (\ref{coll}), $\Delta t$ in (\ref{clock5}) is infinite (or very large in the more realistic case).
Despite the uniform notation ``$\Delta t$'', this large $\Delta t$ in (\ref{clock5})
is totally unrelated to the small $\Delta t$ in (\ref{e2}). These two $\Delta t$'s
refer to two different physical events; the small one is the duration of the box opening,
while the large one, relevant in (\ref{clock5}), is associated with the measurement of energy.

As we have seen in Sec.~\ref{SEC2},
neither Einstein nor Bohr have recognized in 1930 that the measurement of the box 
energy after the emission influences the uncertainty $\Delta E$ of the photon. They both tacitly assumed 
in this thought experiment that $\Delta E$ of photon before the box measurement 
must be the same as $\Delta E$ of photon after the box measurement.
Indeed, in 1930 it seemed absurd to both Einstein and Bohr that some operation performed
on the box could have any influence on the decoupled photon traveling away from the
box with the speed of light.
Nevertheless they were wrong, for the reason which brings us to the next subsection.

\subsection{The role of nonlocality}

The second  important consequence of (\ref{coll})  is {\em nonlocality}. Indeed, similarly to the argument
in the 1935 EPR paper \cite{EPR}, in 1930 Einstein could argue that
measurement of the box after the photon emission cannot influence $\Delta E$ (or any other property) of the
photon, because it would imply that some information from the box to the photon should travel faster than light.
But today we know that Einstein was wrong \cite{bell,merm1,merm2,jord,merm3,laloe,nik-myth,scholarpedia}. 
The measured properties of one object
may be strongly correlated with the measured properties of another object at a large
distance from the first, such that no classical communication (involving only signals 
which don't travel faster than light) between them is possible.
We see that such a nonlocal effect is exactly what happens in the Einstein-Bohr thought
experiment. The measurement
of the box induces the collapse (\ref{coll}), which nonlocally affects the state of the 
photon. 

It is also instructive to compare it with the thought experiment in the EPR paper \cite{EPR},
which is usually considered to be the first example of a nonlocal effect in QM.
For that purpose let us briefly present the main ideas of EPR, in a form somewhat simpler
than in the original paper. 
Instead of time-energy uncertainty relation, the EPR paper deals with the momentum-position
uncertainty relation 
\begin{equation}\label{dpdx}
\Delta p \, \Delta x \gtrsim \hbar . 
\end{equation}
For that purpose, instead of (\ref{Psi}) one deals with the state
\begin{equation}\label{epr}
 |\Psi\rangle =\frac{1}{\sqrt{2}} [|p\rangle |-p\rangle + |-p\rangle |p\rangle ] ,
\end{equation}
where $|p\rangle$ and $|-p\rangle$ are 1-particle momentum eigenstates.
The momentum of each particle can be either $p$ or $-p$, so the 1-particle momentum is uncertain
for each of the particles.
Nevertheless, both terms in (\ref{epr}) have zero total momentum, i.e., $p+(-p)=0$ in the
first term and $(-p)+p=0$ in the second. Thus, the total momentum is not uncertain.

Now let us measure momentum of the first particle and position of the second particle.
For definiteness, suppose that the momentum measurement of the first particle
gives the value $p$. This means that (\ref{epr}) suffers the collapse
 \begin{equation}\label{epr_col}
|\Psi\rangle \rightarrow |p\rangle |-p\rangle .
\end{equation}
But (\ref{epr_col}) determines also the momentum of {\em second} particle, equal to $-p$.
So we know both the momentum of the second particle (through the momentum measurement
of the first particle) and the position of the second particle (through the position measurement 
of the second particle), in contradiction with (\ref{dpdx}). From that contradiction,
EPR concluded that QM was incomplete. 

As well understood today, the correct resolution of the EPR contradiction is not
incompleteness, but nonlocality. Namely, the momentum measurement of the first particle 
affects also the state of the second particle. Thus, the momentum measurement
leading to (\ref{epr_col}) implies that the state of the second particle is $|-p\rangle$,
which has an infinite position-uncertainty $\Delta x$. Consequently, contrary to 
the EPR argument above, one cannot determine
both momentum and position of the second particle. The contradiction obtained by EPR was an 
artefact of the wrong assumption that measurement on one particle cannot 
influence the properties of the other. 

The first experimental verification of quantum nonlocality
has been performed in 1972 by Freedman and Clauser \cite{EPR-test}.
They have measured the linear polarization correlation
of the photons emitted in an atomic cascade of calcium.
Their results were in agreement with predictions of QM
and in contradiction with locality expressed as a variant
of Bell inequality derived by Clauser, Horne, Shimony and Holt
\cite{CHSH}. 
Since then, many other improved experimental verifications
of quantum nonlocality have been performed as well (see, e.g., 
\cite{genovese} for a review), the most famous one
being the experiment performed by Aspect \cite{aspect}.

In essence, we see that the 1935 EPR argument against the momentum-position
uncertainty relation is very similar to the 1930 Einstein argument 
against the energy-time uncertainty relation. They both involve entanglement between
two objects. (The entanglement describes energy correlations (\ref{Psi}) in one case  
and momentum correlations (\ref{epr}) in the other.) 
They both involve an assumption that measurement on one object
cannot influence the properties of the other. They both involve a collapse into a state
in which each of the two objects has a definite value of the relevant observable
(energy in (\ref{coll}) and momentum in (\ref{epr_col})). And finally, 
to correctly reply to these two arguments against uncertainty relations, 
in both cases one needs to invoke quantum nonlocality. 

Therefore, contrary to the wide belief, the famous EPR paper was not the first 
serious challenge of QM that required quantum nonlocality for a correct resolution.
Einstein presented such a challenge already in 1930, five years before the EPR paper.

\section{Conclusion}
\label{SEC4}

In 1930 Einstein presented a challenge for QM, 
by considering a thought experiment involving a measurement of energy of the 
box which emitted a photon. Bohr attempted to resolve this challenge
by appealing to Einstein's general theory of relativity. Even though
it is often presented in the literature as a triumph of Bohr over Einstein,
in this paper we have revisited this thought experiment 
from a modern point of view and found out that Bohr's resolution was not correct.
Instead, the correct resolution involves quantum nonlocality
originating from the entanglement between box and photon.

Since this 1930 Einstein's challenge of QM
has been presented five years before the EPR challenge of QM, 
and since these two challenges are very similar in essence, 
and since the correct resolution of both involves quantum nonlocality 
as a crucial ingredient,
we conclude that it is incorrect to think of the EPR paper
as the first historical example of a quantum system which cannot be understood correctly without invoking quantum nonlocality.

\section*{Acknowledgments}

The author is grateful to V. Hnizdo and H. D. Zeh  
for valuable comments on the manuscript and 
for drawing attention to some relevant references.
This work was supported by the Ministry of Science of the
Republic of Croatia under Contract No.~098-0982930-2864.

\end{document}